# A record of atmospheric $^{210}$Pb accumulation in the industrial city

E.A. Buraeva, V.S. Malyshevsky[1], V.V. Stasov, L.V. Zorina

*Southern Federal University, 344090, Rostov-on-Don, Russia*

B.I. Shramenko

*National Science Center «Kharkov Physical-Technical Institute», 61108, Kharkov, Ukraine*

**Abstract**

The deposition flux of total atmospheric $^{210}$Pb in the industrial city Rostov-on-Don, Russia from 2002 to 2010 has been measured. The variations in annual $^{210}$Pb deposition flux appear to be mainly correlated with the number of rains and significant amount of anthropogenic $^{210}$Pb, polluted into the surface layer of air in the home-heating period. The average $^{210}$Pb deposition is 1.75 mBq/m$^3$. Several meteorological parameters which are strongly associated with the fluctuations of concentrations of $^{210}$Pb are identified. These results are useful to provide typical information on the atmosphere radioactivity in an industrial city.

## 1. Introduction

Natural radionuclide of terrestrial origin $^{210}$Pb is decay products of $^{238}$U. The half-lives of this radionuclide is 22.3 years. The $^{210}$Pb appears in the atmosphere from the decay of inert gas $^{222}$Rn. The higher levels of radon in the air is characteristic of areas with radioactive anomalies in uranium deposits, as well as in areas of tectonic faults. Thus the $^{222}$Rn (and as a result $^{210}$Pb) may be either of natural and technogenic origin (for instance, due to various processes with the mineral raw materials). The purpose of this paper is to determine the characteristics of the climate effects, weather, and possibly man-made factors on the variation of $^{210}$Pb in aerosols and precipitation in the surface air of industrial city Rostov-on-Don, Russia (moderate latitudes and temperate arid climate. Latitude: 47° 14.1834' N, Longitude: 39° 42.8334' E).

The long-term monitoring of radioactivity in air provides useful information about the radiation in the environment due to natural and man-made sources and helps to evaluate their impact on man. Monitoring of the $^{210}$Pb behavior in the atmosphere is carried out in different laboratories. Here are just some of the earlier studies. For instance, concentration $^{210}$Pb in air was continuously monitored, using a high-volume air sampler and a high-resolution gamma-ray spectrometer, at the University of Málaga (Duenas C., et.al. 2009). Studies of radionuclide activities in aerosol particles have been carried out by Duenas C. et.al. (2011). The $^{210}$Pb deposition over the North Sea was investigated by Beks J.P. et. al. (1998). Activity concentrations of the long-lived natural radionuclide $^{210}$Pb in surface air were measured by Hotzl H. et.al. (1987) twice monthly at a semi-rural location 10 km north of Munich (FRG) for the time interval 1983-1985. The $^7$Be to $^{210}$Pb concentration ratios in ground level air on two monitoring stations in Belgrade area were determined from 1996 to 2001 by Todorovic D. et.al (2005).

---

[1] Corresponding author. E-mail address: vsmalyshevsky@sfedu.ru

According to Todorovic D. the $^7$Be to $^{210}$Pb ratios were in the range of 1.7–12.7, with summer maxims and winter minimums. There is a strong correlation between the seasonal radionuclides concentrations and the seasonal average of temperature; it is quite likely this is one of the master variables that control the amount of radionuclides concentrations in air (Azahra M., et. al. 2004).

The results presented in this paper are a part of research on atmospheric radioactivity in the south of Russia (Buraeva E.A. et. al. 2007).

## 2. Experimental

Measurements of the $^{210}$Pb content in aerosols (1 per week) and precipitation (1 per month) are performed at the aspiration station of the Southern Federal University (Rostov-on-Don, Russia) in 2002–2010 as part of the monitoring of the radioactivity of the atmospheric layer at the ground in Rostov-on-Don (47°14′ NL; 39°42′ EL). The location of the station at temperate latitudes with a temperate continental climate and comparatively low precipitation imparts special significance to the systematic monitoring of $^{210}$Pb in the atmosphere.

A ventilation setup with a filter consisting of FPP-15-1.7 Petryanov fabric with total area 0.56 m$^2$ and a liquid Lambrecht micromanometer were used to obtain the samples. According to the measurements, the air flow rates were approximately 630 m$^3$/h initially ("fresh" filter) and 510 m$^3$/h after 7 days of exposure. The exposed filter was air dried and pressed into 35 mm in diameter and 10–30 mm high pellets. Three or four days after the filter was removed, the γ-ray spectrum was measured in 12–24 h with a Ge(Li) or HPGe detector of the low-background setup. The dust content in air was found according to the mass difference between the exposed and clean filter.

## 3. Results and discussion

The $^{210}$Pb volume activity in the lower atmosphere of industrial city Rostov-on-Don (Russia) was measured during 2002-2010. In general, the activity of $^{210}$Pb in the lower atmosphere varies within a wide range (0,12 - 9,05) mBq/m$^3$, with the average for the whole period of observation 1.75 mBq/m$^3$ (see Fig.1 and Table 1).

Obviously, regardless of the origin of $^{210}$Pb (natural or man-made) its content in atmospheric aerosols decreases with the increase in rainfall due to more intensive leaching. Additionally for the natural $^{210}$Pb decreases due to reducing the rate of radon exhalation from the moist soil as well. The maximum values of the activity 2.2 - 3.3 mBq/m$^3$ achieved with a combination of relatively high temperatures 16 - 26 $^0$C and the relatively small amount of precipitation 29 - 43 mm/month (Table 2).

Seasonal dependence of the $^{210}$Pb activity is shown in Fig. 2. To illustrate the overall picture of the relationship of the individual parameters that characterize the radioactivity of the atmosphere, we present their seasonal dependence, averaged over the whole observation period (Fig.3). It is seen that in the winter months with minimal amounts of precipitation and minimum temperatures the $^{210}$Pb activity in aerosols reach the highest values. This can only be associated with a significant amount of anthropogenic $^{210}$Pb, polluted into the surface layer of air in the home-heating period. High values of $^{210}$Pb activity of anthropogenic aerosols are accompanied by stability of the atmosphere in the winter. A slight increase in activity in the winter after the 2006 is associated with colder winters in these years and as a consequence a great burning of fossil fuels. The low $^{210}$Pb activity values in aerosols in this period (for instance in March) associated with the turbulent mixing in certain months.

For the spring-summer period the behavior of the $^{210}$Pb activity are explained by the restructuring of the spring atmosphere and the maximum amount of precipitation in June. This maximum rainfall separates spring and summer maximum of 210Pb activity for two (April and September) and deposition density of $^{210}$Pb on the Earth's surface in June.

**Acknowledgement**

This work is supported by Federal Program of the Russian Ministry of Science and Education "Scientific and scientific-pedagogical personnel of innovative Russia" (grant number 14.A18.21.0633) and Russian Foundation for Basic Research and by National Academy of Sciences of Ukraine grant number 12-08-90401-Ukr_a.

# Tables

Table 1. Volume activity of $^{210}$Pb in surface air.

| Year | Volume activity, mBq/m$^3$ | | |
|---|---|---|---|
|  | Max | Min | Mean |
| 2002 | 4,96 | 0,32 | 1,41 |
| 2003 | 7,40 | 0,12 | 1,55 |
| 2004 | 3,40 | 0,20 | 1,26 |
| 2005 | 3,28 | 0,48 | 1,40 |
| 2006 | 8,84 | 0,68 | 2,19 |
| 2007 | 9,05 | 0,15 | 1,85 |
| 2008 | 8,44 | 0,71 | 2,05 |
| 2009 | 7,90 | 1,00 | 2,27 |
| 2010 | 5,41 | 0,94 | 2,25 |

Table 2. Volume activity of $^{210}$Pb and meteorological parameters averaged over the year

| Year | Activity, mBq/m$^3$ | Precipitation, mm | Temperature, $^0$C | Wind speed, m/s | Relative humidity, % | Atmospheric pressure, mm Hg |
|---|---|---|---|---|---|---|
| 2002 | 1,41 | 38,5 | 10,0 | 1,8 | 65,9 | 756,1 |
| 2003 | 1,55 | 50,7 | 8,3 | 2,0 | 71,5 | 756,7 |
| 2004 | 1,26 | 25,3 | 9,6 | 1,8 | 74,6 | 755,8 |
| 2005 | 1,40 | 32,7 | 10,3 | 1,9 | 72,3 | 756,2 |
| 2006 | 2,19 | 38,2 | 10,6 | 3,2 | 69,4 | 755,7 |
| 2007 | 1,85 | 37,6 | 10,9 | 4,3 | 65,1 | 755,6 |
| 2008 | 2,05 | 27,6 | 11,1 | 4,2 | 69,4 | 756,1 |
| 2009 | 2,27 | 27,3 | 11,0 | 4,2 | 71,6 | 755,2 |
| 2010 | 2,25 | 25,3 | 9,7 | 5,6 | 71,3 | 755,3 |

**Captions**

Fig. 1. Volume activity of the $^{210}$Pb.

Fig. 2. Volume activity of $^{210}$Pb in surface air from January 2002 since December 2010.

Fig. 3. Mean monthly atmospheric volume activity of $^{210}$Pb averaged over the period 2002-2010.

**Figures**

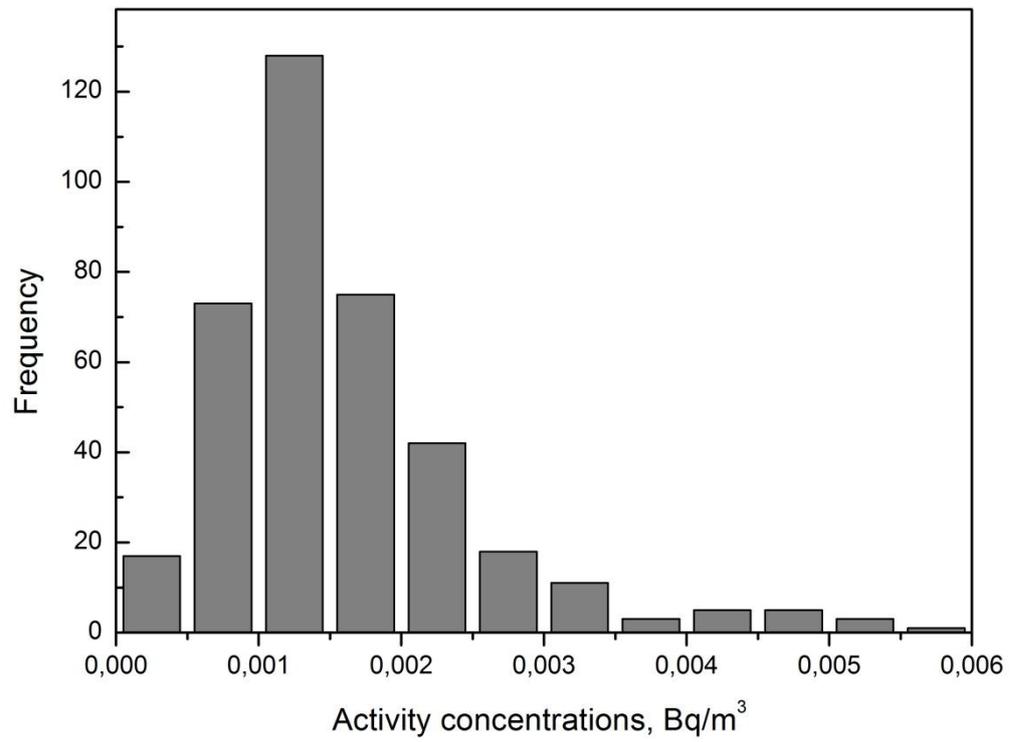

**Fig. 1.**

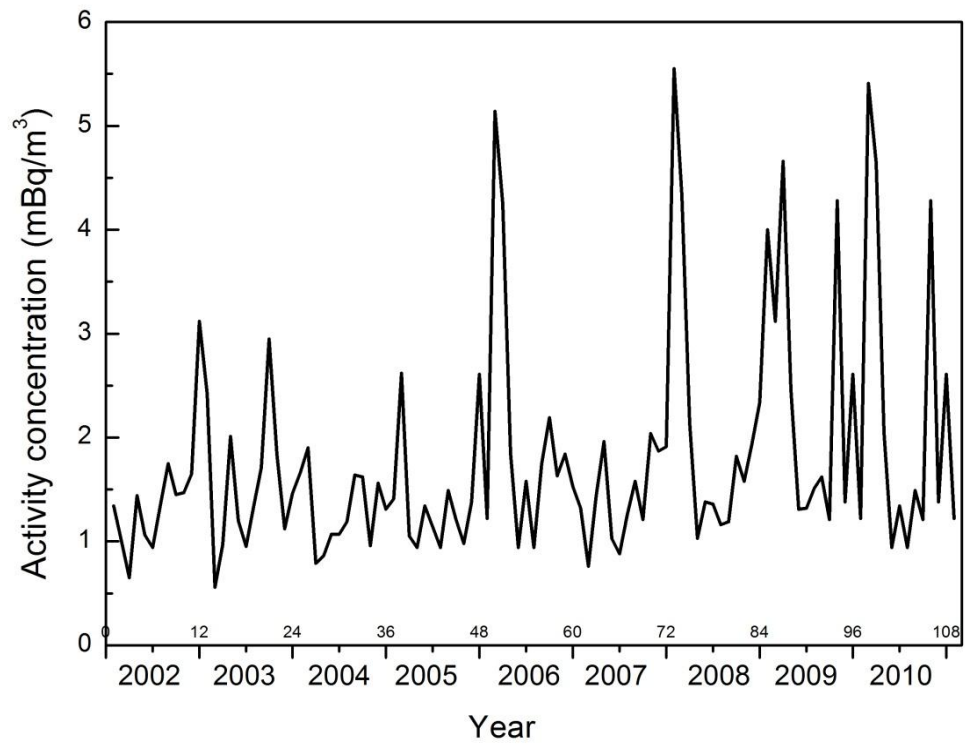

**Fig. 2.**

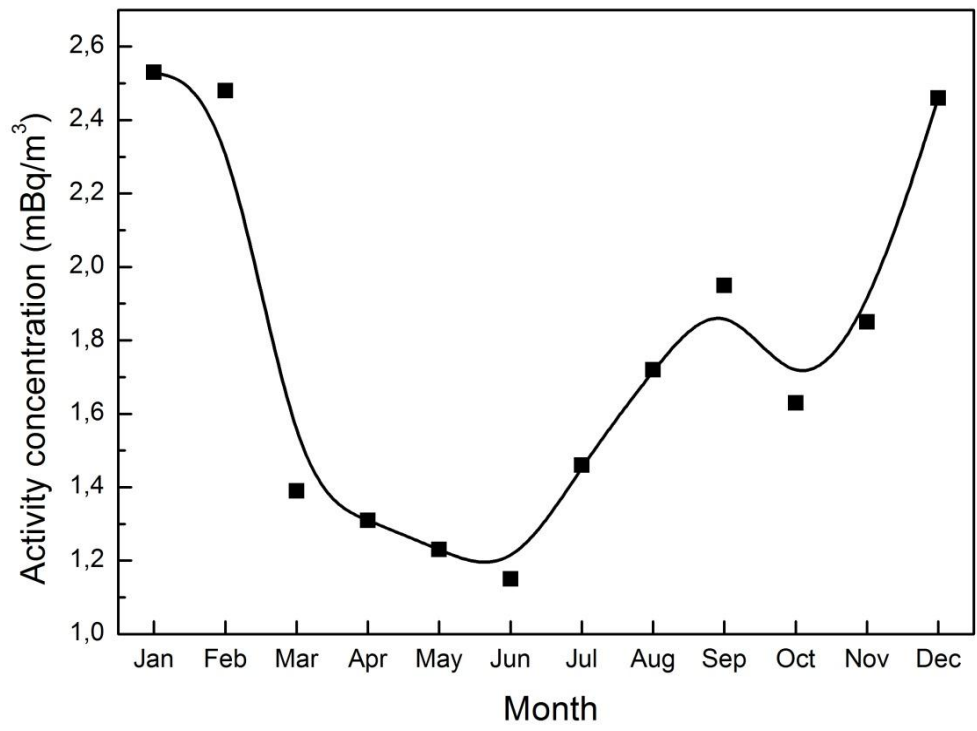

**Fig. 3.**